# Using Lexical Features for Malicious URL Detection - A Machine Learning Approach

*Apoorva Joshi*
FireEye Inc.

*Levi Lloyd*
Lawrence Livermore National Laboratory

*Paul Westin*
FireEye Inc.

*Srini Seethapathy*
FireEye Inc.

**ABSTRACT**

Malicious websites are responsible for a majority of the cyber-attacks and scams today. Malicious URLs are delivered to unsuspecting users via email, text messages, pop-ups or advertisements. Clicking on or crawling such URLs can result in compromised email accounts, launching of phishing campaigns, download of malware, spyware and ransomware that may result in severe monetary losses. Hence, it is imperative to detect and block these threats effectively. The standard and fastest way to identify malicious URLs is by comparing URLs against blacklists (repositories of known malicious websites, domains and hosts) [1]. However, blacklists are never exhaustive and lack the ability to detect newly generated URLs. Recognizing the extant need and shortcomings of blacklists-based methods, a machine learning based ensemble classification approach is proposed herein to combat the above threat. The approach uses static lexical features extracted from the URL string, with the underlying assumption that the distribution of these features is different for malicious and benign URLs. The use of such static features is safer and faster since it does not involve execution [2]. The model has been deployed in the FireEye Advanced URL Detection Engine (FAUDE), resulting in a significant increase in malicious URL detections.

**INTRODUCTION**

The Internet has become a breeding ground for various kinds of fraud as increasing number of individuals, services and businesses get online for fulfilling their respective requirements. It is difficult for Internet users to be aware of the latest threats, scams, malware and other forms of malicious activity. Therefore, extensive research is being carried out for developing reliable methods to protect Internet users from unanticipated harm.

Machine learning algorithms have been proven to be useful in detecting maliciousness [3]. Previous efforts have built static classifiers using URL lexical features, host information, network traffic and other schemes. Many of these approaches use a Bag-of-Words approach which results in very large feature vectors. As retrieving host information involves lookups on remote servers, a significant amount of latency is inherent in these approaches and prevents their use in lightweight real-time systems which are designed to produce fast verdicts. Prior to this work, the potential of a fully lexical detection approach for production level workloads had not yet been successfully explored.

This paper investigates the use of an ensemble machine learning approach in detecting malicious URLs delivered in emails. A variety of approaches have been used previously to tackle the problem of malicious URL detection. Blacklists are a common and classic method for detecting malicious URLs due to their simplicity. Heuristic based approaches are an extension of blacklists wherein a blacklist of signatures of common attacks is maintained. The major drawback of both blacklist and heuristic based methods is their inability to adapt to new threats [2]. Machine learning and deep learning models which use a combination of static information like lexical features from the URL string, host information, and sometimes even HTML and JavaScript content have also been explored for URL classification [4-7].

We present a Random Forest model for URL classification using purely static lexical features extracted from the URL string. The methodologies and algorithms used in this paper can also be extended to URLs delivered via other platforms like text messages, advertisements etc.

**METHODOLOGY**

The major steps associated with the Crisp-DM Methodology used to build the classification model are outlined in the following diagram [8]:

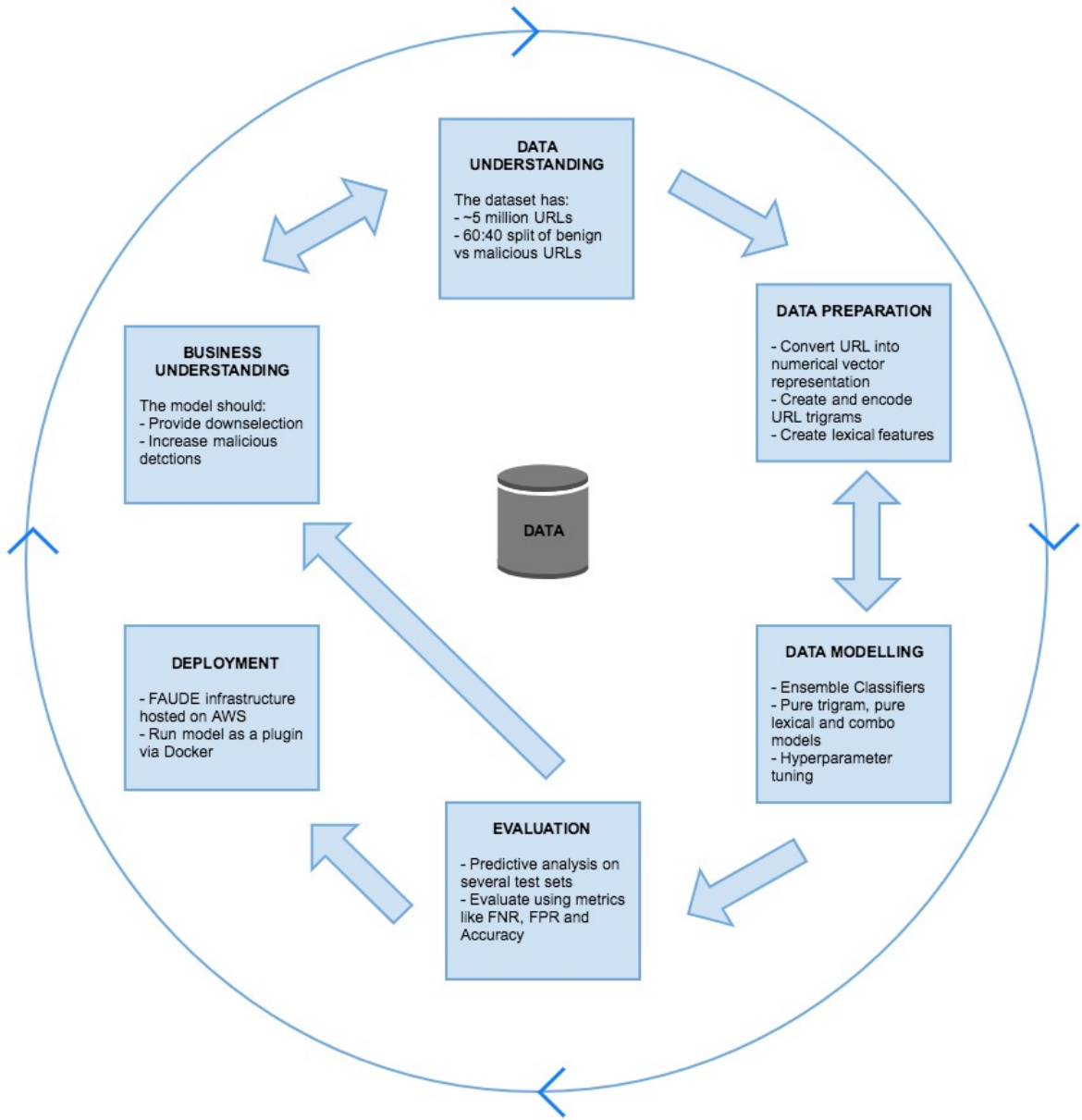

**Figure 1: Main steps involved in Crisp-DM Methodology**

A detailed explanation of each of the major phases in the Crisp-DM process model as applied to this project, is as follows:

**Business Understanding**
The objective was to integrate the classification model into the FireEye Advanced URL Detection Engine (FAUDE), either as a means of down-selection, detection or both. As a result, there are performance requirements from the model like reasonable model size, low latency and low False Negative Rate (FNR) that need to be met without compromising greatly on the False Positive Rate (FPR) and Accuracy.

**Data Understanding**
The dataset used for training the classification model consisted of ~5 million URLs collected from various sources, including online repositories like Openphish, Alexa whitelists and internal FireEye sources, to get a varied collection of malicious and benign URLs. A 60-40 split was maintained between benign and malicious URLs so as to have a good representation of malicious and benign URLs in the dataset.

**Data Preparation**
The first step in building the classification model was to extract the features for the model. Machine learning algorithms can only work with numerical inputs. Hence, the URL strings need to be encoded into meaningful numerical vectors. After analysis of several URLs, we found 23 different lexical features that could be used to differentiate between malicious and benign URLs. The lexical features were combined with 1000 trigram-based features to produce 1023-long numerical vectors to represent the URLs. This paper uses NLTK (a popular NLP Python package) to create URL trigrams, mmh3 (a Murmur Hash Python package) to encode the trigrams, and urrlib (a Python library for parsing URLs) to parse the URLs to obtain the lexical features from them. The 23 lexical features are listed in Table 1.

Correlation and scatter matrices were used to find the interdependencies between the features, in order to extract only the most significant ones. It was observed that many of the length-based lexical features were highly correlated. For every pair of features with a correlation coefficient greater than 0.75, only one of the features was retained and the other discarded.

As also noted in previous studies [2], it was observed that the lexical structure of malicious URL strings is significantly different from that of benign URL strings. For example, malicious URLs on an average have several levels of sub-domains, special characters in the URL path. Usually, their domain names resemble whitelist domains, but with minor changes, higher average domain lengths- patterns which are not usually seen in benign URLs. Hence, lexical features have the potential to distinguish between malicious and benign URLs and were used to build features for the classification.

**Data Modelling**
URL strings tend to be very unstructured and noisy. Hence, it was imperative to choose a classification algorithm that is not too sensitive to fluctuations in the training data. Another important research direction was also to build an interpretable model so as to understand and validate what patterns in URLs make them malicious. This is particularly difficult when deep learning models are used, which often behave like black boxes. Comparison of different machine learning models showed that bagging algorithms (Random Forest in this case) were a good fit for the task since they average out multiple learners trained on different parts of the training data, with

the goal of reducing variance. Low latency was ensured since lexical features are direct derivatives of the URL string and thus very efficient to compute.

Random Forest with Decision Tree estimators was chosen as the model for classification. The hyperparameters of the Random Forest were tuned so as to maintain a reasonable trade-off between predictive power of the model and model size.

**Table 1: List of lexical features**

| URL Component | Lexical Feature |
| --- | --- |
| URL | Length |
| URL | Number of semicolons, underscores, question marks, equals, ampersands |
| URL | Digit to letter ratio |
| Top level domain | Presence in suspicious list |
| Primary domain | Contains IP |
| Primary domain | Length |
| Primary domain | Number of digits |
| Primary domain | Number of non-alphanumeric characters |
| Primary domain | Number of hyphens |
| Primary domain | Number of @s |
| Primary domain | Presence in top 100 Alexa domains |
| Subdomain | Number of dots |
| Subdomain | Number of subdomains |
| Path | Number of '//' |
| Path | Number of subdirectories |
| Path | Presence of '%20' in path |
| Path | Presence of uppercase directories |
| Path | Presence of single character directories |
| Path | Number of special characters |
| Path | Number of zeroes |
| Path | Ratio of uppercase to lowercase characters |
| Parameters | Length |
| Query | Number of queries |

**Evaluation**

The results of the Random Forest model were compared against simple machine learning classifiers such as Naïve Bayes, SVM and Logistic Regression, as well as other ensemble classifiers like AdaBoost and Gradient Boost. Table 2 summarizes various metrics like Accuracy, Area under the ROC curve (AUC) and FNR, which were used to compare between the three algorithms.

**Table 2: Comparison of ensemble classifiers**

| Algorithm | Accuracy (%) | AUC | FNR (%) |
|---|---|---|---|
| Random Forest | 92 | 0.99 | 0.38 |
| Gradient Boost | 90 | 0.92 | 9 |
| AdaBoost | 90 | 0.9 | 10 |
| Logistic Regression | 87 | 0.96 | 4.75 |
| Naïve Bayes | 70 | 0.74 | 10.38 |

The above comparison shows that Random Forest was the best choice for the classification, given that it has the highest accuracy and lowest FNR among the models compared. Table 3 compares the performance of different feature extraction strategies that were used along with Random Forest for classification.

**Table 3: Comparison of feature extraction strategies**

| Number of trigrams | Number of lexical features | FPR (%) | FNR (%) | Accuracy (%) |
|---|---|---|---|---|
| 1000 | 0 | 29.8 | 0.4 | 85 |
| 1000 | 23 | 16.8 | 0.38 | 92 |
| 300 | 23 | 12.5 | 0.93 | 93 |
| 100 | 23 | 11.5 | 1.09 | 94 |
| 0 | 23 | 8.15 | 1.1 | 95 |

The above comparison shows that a combination of 1000 trigram-based features along with lexical features best achieves the requirements from the classification model i.e. low FNR without greatly compromising on the FPR and good accuracy. Hyperparameter tuning of the Random Forest involved tuning mainly the maximum depth of trees of the Random Forest, as summarized in Table 4. It was observed that tuning hyperparameters like the minimum samples required to split at a node and minimum samples required at a leaf node did not have a substantial effect on the accuracy or FNR for this model.

**Table 4: Hyperparameter tuning summary**

| Max depth | FNR (%) | Accuracy (%) |
|---|---|---|
| 5 | 1.13 | 72 |
| 10 | 0.81 | 81 |
| 15 | 0.48 | 88 |
| 20 | 0.38 | 92 |
| 27 | 0.73 | 94 |
| 30 | 0.75 | 95 |

The above statistics show that a maximum tree depth of 20 gave the best tradeoff between FNR and accuracy.

**Deployment**

The FAUDE architecture is hosted on AWS and the classification model runs in a docker container, as a plugin (called RF Selector) for FAUDE. The plugin has been deployed as a level of down-selection between the fast (Fastpath) and slow (Slowpath) URL analysis components of FAUDE. URLs marked as benign by the model are rejected while those marked as malicious are selected for further analysis. Since its deployment in the production environment, there has also been a 22% increase in malicious detections on FAUDE. Fig.2 is a high-level representation of how RF Selector fits into the FAUDE pipeline:

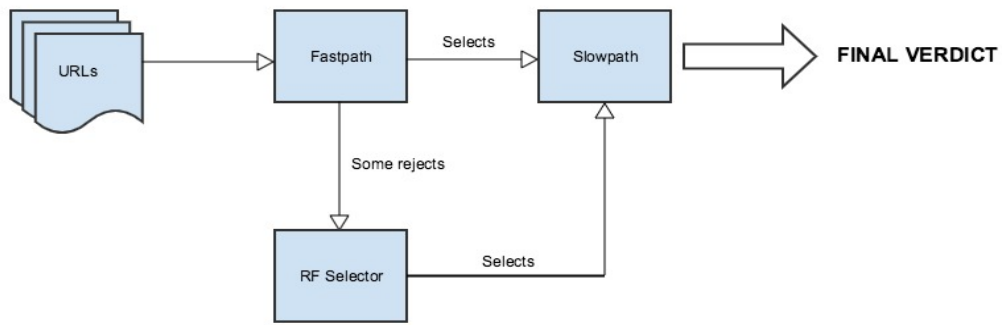

Figure 2: RF Selector in FAUDE

**CONCLUSION**

This paper proposes a static lexical feature-based Random Forest Classification approach to classify malicious and benign URLs. The results obtained from the study show noteworthy evidence that purely lexical approaches can be used in lightweight systems to generate fast real-time verdicts for URLs or websites on the Internet.